# High-pressure synthesis, crystal and electronic structures of a new scandium tungstate, $Sc_{0.67}WO_4$

Tamas Varga,[1*] J. F. Mitchell,[2] Jun Wang,[3**] Lindsay G. Arnold,[3] Brian H. Toby,[3] and Christos D. Malliakas[2,4]

[1]Environmental and Molecular Sciences Laboratory, Richland, WA 99354, USA

[2]Materials Science Division, Argonne National Laboratory, Argonne, IL 60439, USA

[3]Advanced Photon Source, Argonne National Laboratory, Argonne, IL 60439, USA

[4]Department of Chemistry, Northwestern University, Evanston, IL  60208, USA

*Corresponding author.
E-mail: tamas.varga@pnl.gov; Fax: +1-509-371-6242
**Current address: National Synchrotron Light Source, Brookhaven National Laboratory, Upton, NY 11973-5000

**Abstract**

Negative thermal expansion (NTE) materials possess a low-density, open structure which can respond to high pressure conditions, leading to new compounds and/or different physical properties. Here we report that one such NTE material—white, insulating, orthorhombic $Sc_2W_3O_{12}$—transforms into a black compound when treated at 4 GPa and 1400 °C. The high pressure phase, $Sc_{0.67}WO_4$, crystallizes in a defect-rich wolframite-type structure, a dense, monoclinic structure (space group $P2/c$) containing 1-D chains of edge-sharing $WO_6$ octahedra. The chemical bonding of $Sc_{0.67}WO_4$ vis-à-vis the ambient pressure $Sc_2W_3O_{12}$ phase can be understood on the basis of the Sc defect structure. Magnetic susceptibility, resistivity, thermoelectric power and IR spectroscopic measurements reveal that $Sc_{0.67}WO_4$ is a paramagnet whose conductivity is that of a metal in the presence of weak localization and electron-electron interactions. Oxygen vacancies are suggested as a potential mechanism for generating the carriers in this defective wolframite material.

## 1. Introduction

Negative thermal expansion (NTE) compounds are opportune candidates for seeking out new compounds at high pressure. This is due to the combination of their flexible, low-density framework structure, [1,2] and the presence of lattice modes that soften on compression. [3-6] The $A_2M_3O_{12}$ family of NTE materials [7,8] is one such group of oxides that has been studied extensively under high pressures and high temperatures. [9-21] Prototype material of the above family, orthorhombic $Sc_2W_3O_{12}$ - a white insulating powder - has been known to transform to a monoclinic structure on compression, [16,17,21] and to undergo pressure-induced amorphization under more extreme pressure conditions. [9] To date, no experimental data for simultaneous high-temperature (>400 °C), [22,23] and high-pressure processing of $Sc_2W_3O_{12}$ have been published.

Here we report that orthorhombic $Sc_2W_3O_{12}$ transforms into a new metallic compound, $Sc_{0.67}WO_4$, when treated at 4 GPa and 1400 °C. $Sc_{0.67}WO_4$ crystallizes in a highly defective wolframite structure, a dense, monoclinic structure (space group *P2/c*) characterized by 1-D chains of edge-sharing $WO_6$ octahedra isostructural to that of multiferroic $MnWO_4$. [24-28] To our knowledge, this new scandium tungstate, $Sc_{0.67}WO_4$, is the first reported compound crystallizing in the wolframite-type structure with trivalent A-site cations. $Sc_{0.67}WO_4$ is an n-type conductor that shows Pauli-type magnetic susceptibility, a negative thermopower linear in T, and nearly linearly increasing electrical conductivity in the range 3-300 K with a finite T → 0 intercept. These data indicate that $Sc_{0.67}WO_4$ is a poor metal with a finite density of states at $E_f$ with conductivity akin to that of highly defective amorphous semiconductors. We suggest that a small concentration of O defects introduces carriers into an impurity state-derived band, and that the mobility of these carriers may be influenced by the highly defective structure of $Sc_{0.67}WO_4$. More generally, the present study confirms the expectation that NTE compounds can serve as precursors to high-pressure synthesis of new and interesting materials via modification of bonding and electronic structure under extreme conditions.

## 2. Experimental

*2.1. Sample preparation*

$Sc_2W_3O_{12}$ powder was prepared from $Sc_2O_3$ (99.9%, Strem Chemicals, Newburyport, MA) and $WO_3$ (99.9%, Aldrich, Milwaukee, WI). Stoichiometric amounts of the two oxides were thoroughly mixed and ground. The mixture was initially heated at 1000 °C for 5 hours and, after regrinding, it was heated at 1200 °C for an additional 12 hours in air. Powder XRD (PANalytical X'Pert Pro, Co $K_\alpha$ radiation) showed this precursor to be single phase. Polycrystalline samples of $Sc_{0.67}WO_4$ were prepared from $Sc_2W_3O_{12}$ powder by high-pressure synthesis in a cubic multianvil press (Rockland Research Co, West Nyack, NY). The high-pressure apparatus is described elsewhere. [29] $Sc_2W_3O_{12}$ powder was loaded into a sealed Pt capsule and compressed to ~4 GPa. The Pt cells containing the samples were resistively heated to ~1400 °C, held for 1 hour, then quenched to room temperature before releasing the pressure slowly. The sintered sample pellet was dense and black. The face of the pellet was polished to remove any potential contaminations. Two samples, S1 and S2, were prepared in this way. Sample mass in each case was ~ 0.1 g.

*2.2 Sample characterization*

Synchrotron powder x-ray diffraction (SXRD) data were collected on sample S1 at a wavelength of 0.401555 Å at the 11-BM-B beamline of the Advanced Photon Source at Argonne National Laboratory. Structural determination was carried out by Rietveld refinement [30] using the software package GSAS+EXPGUI. [31,32] For the Rietveld analysis, atomic coordinates, independent isotropic temperature factors for each site, 16 terms determining a Shifted Chebyschev fit to the background, three Gaussian broadening terms, one Lorentzian term, and nine microstrain broadening terms in the peak shape model were refined for the main phase. Structural parameters and agreement indices from all these refinements are available as Supplementary Material. Semiquantitative microprobe elemental analysis was performed on several $Sc_{0.67}WO_4$ crystals using a Hitachi S-2700 scanning electron microscope (SEM) equipped with a Noran energy-dispersive spectroscopy (EDS) detector. Magnetization and specific heat measurements were performed on a Quantum Design Magnetic Property Measurement System (MPMS). Four-probe electrical transport data were measured on a home-built system. Thermoelectric power of the sample was measured under vacuum between 305-600 K in 5 K steps, on a Seebeck coefficient measurement system from MMR Technologies using 100 mW power. Thermogravimetric analysis (TGA) was carried out on a Mettler-Toledo TGA851 instrument in both nitrogen and oxygen atmospheres. For the determination of the band gap, a spectrum was recorded in the Mid-IR region (4000 – 400 $cm^{-1}$) at room temperature with a Nicolet 6700 FT-IR Spectrometer equipped with a diffuse reflectance collector from Spectra-Tech Inc. The optical band gap was determined using the Kubelka-Munk theory. [33-35]

## 3. Results

A comparison of the powder x-ray diffraction patterns of the starting material orthorhombic $Sc_2W_3O_{12}$, and the end product (Sample S1) containing the new phase is shown in Fig. 1; data from S2 were similar. Indexing the strong lines led to a monoclinic cell, $a$ = 4.80 Å x $b$ = 5.76 Å x $c$ = 4.99 Å, β = 91.18°, in space group $P2/c$). Some impurity phases are present, including some residual starting orthorhombic $Sc_2W_3O_{12}$, an orthorhombic non-stoichiometric perovskite phase (possibly having composition $Sc_{0.3}W_{0.85}O_3$, in analogy to the previously reported $Er_{0.05}Ta_{0.15}W_{0.85}O_3$ [36]), and some other unidentified impurities, in minor amounts (see Fig. 1). A search of the ICDD-JCPDS database [37] revealed a good match of the lattice constants to the wolframite-type structure of $MnWO_4$. [24] To refine the structure of the new phase, we collected SXRD data on the sample and analyzed these data with the Rietveld method, using the published crystal structure of $MnWO_4$ [24] as the starting model. The lattice parameters, compared to those of $MnWO_4$, measured Sc/W fractional occupancies and the fractions of crystalline phases present obtained from the Rietveld fit to the SXRD data are tabulated in Table 1. A representative Rietveld profile fit is shown in Fig. 2. Fractional coordinates, interatomic distances and bond angles are listed in the Supplementary Material.

$Sc_{0.67}WO_4$, crystallizes in the wolframite structure, a dense, monoclinic structure (space group $P2/c$) containing 1-D chains of edge-sharing $WO_6$ octahedra (Fig. 3). Neglecting the few weak unindexed lines, the sample contained 86.96(10) wt% $Sc_{0.67}WO_4$, 3.55(6) wt% $Sc_2W_3O_{12}$, and 9.49(6) wt% of a perovskite phase tentatively assigned as $Sc_{0.3}W_{0.85}O_3$. It should be noted that this putative $Sc_{0.3}W_{0.85}O_3$ composition is also a potentially new phase, to the best of our knowledge. However, in the absence of a definitive structural/compositional characterization at this point, this finding remains speculative.

The temperature dependence of the field-cooled (H = 0.1 kOe) DC susceptibility of Sc$_{0.67}$WO$_4$ (sample S1) is consistent with temperature-independent Pauli paramagnetism superimposed with local moments at lower temperatures. Across the 2-300 K temperature range, the data shown in Fig. 4(a) obey a modified Curie-Weiss law of the form, $\chi = \chi_{Pauli} + \chi_{core} + \chi_{vv} + C/(T-\theta_W)$, where $\chi_{Pauli}$ is a temperature independent Pauli paramagnetic (TIP) term, $\chi_{core}$ (-65 x 10$^{-6}$ emu/mol) [38] is the diamagnetic core correction, and $\chi_{vv}$ (35 x 10$^{-6}$ emu/mol), estimated from data reported for WO$_3$ [39] is the temperature-independent Van Vleck contribution, and C and $\theta_W$ are the Curie constant and Weiss temperature, respectively. The fit yields $\theta_W$ = -0.33(1) K and an effective moment, p$_{eff}$ = 0.15 $\mu_B$/f.u. and $\chi_{Pauli}$ = 4.4 x 10$^{-4}$ emu/mol. We note that the core and Van Vleck corrections are <10% of the total, implying that the residual TIP arises from carriers at E$_f$. The thermoelectric power of the sample above 300 K (Fig. 5 (a)) is small and negative with a linear T-dependence, consistent with a metallic n-type density of states. Fig. 5(b) shows that the electrical conductivity, $\sigma$, increases linearly with T in the range 30-300 K. Attempts to fit the data to various activated models (e.g., simple exponential, variable-range hopping, adiabatic small polaron, etc.) were unsuccessful. The data reveal that Sc$_{0.67}$WO$_4$ has a finite T → 0 conductivity, albeit that of a poor metal. Below 30 K, the conductivity can be fit by introducing an additional T$^{1/2}$ term as discussed below. The band gap measured by FTIR is ~0.15 eV, far smaller than that found in related wolframite tungstate compounds. [40-42] Thermogravimetric analysis revealed that the sample (S1) is stable up to 600 °C when heated in an oxygen atmosphere; a mass loss of 1.6% was observed. When heated to 400 °C in nitrogen, a much smaller mass loss of 0.05% was observed.

## 4. Discussion

*4.1. Structure*

The monoclinic wolframite structure, and the related tetragonal scheelite structure are known for many metal (alkaline earth, transition and rare earth) tungstates and molybdates with the $ABO_4$ general formula, mainly for A=2+ cations. [28] Molybdates of trivalent metals Eu, Y, and Gd-Lu with a tetragonal scheelite-type structure were first reported by Banks *et al*., [43] and specifically a disordered scheelite-type $Eu_{0.67}MoO_4$ compound containing Eu(III) and Mo(VI) was made, [43] which appeared to be similar to a $La_2(MoO_4)_3$ phase previously suggested by Jeitschko. [44] However, no monoclinic $ABO_4$ phase with A being a 3+ cation and B being Mo or W has been reported to our knowledge.

Prompted by the black color of the sample, we considered the possibility of reduction of W from +6 ($d^0$) to +5 ($d^1$) during the high pressure, high temperature reaction. Considering the wolframite structure, this leads to a hypothetical "$ScWO_4$" composition. Of relevance in this case is the report by Doumerc *et al*. on the synthesis and structure of $AlWO_4$, [45] a high-pressure phase crystallizing in a monoclinic *C*2/*m* structure, and containing W(V) stabilized by W-W pairs. However, "$ScWO_4$" would be Sc-rich relative to the starting material $Sc_2W_3O_{12}$ and no Sc-rich phases were identified in the laboratory powder x-ray diffraction pattern. Prompted by EDS analysis, which gave an average

Sc:W elemental ratio of 2:3, similar to that of the starting material, we collected SXRD data on the sample (S1) and carried out Rietveld analysis using the $MnWO_4$ structure as a starting model (see Table 1 and Fig. 2). Significantly, refinement of the site occupancies unambiguously demonstrates that the Sc site is only 2/3 occupied, giving an $Sc_{0.67}WO_4$ composition, that of the nominal starting composition and consistent with the EDS analysis. No evidence (supercell reflections) indicating any small multiples of the original unit cell was found in the SXRD pattern; detailed electron diffraction experiments would be desirable to corroborate this conclusion.

The structure of $Sc_{0.67}WO_4$ consists of edge-sharing $WO_6$ and $ScO_6$ octahedra. The increase in the coordination of tungsten is consistent with what is generally seen on transformations into higher-density high-pressure phases. As expected, $Sc_{0.67}WO_4$ (volume/formula unit: 68.930(1) $Å^3$) is much denser than the starting phase $Sc_2W_3O_{12}$ (V/f.u.=308.49/3 = 102.83 $Å^3$/W). This corresponds to a ~33 % reduction in volume. The local structure of edge-sharing $WO_6$ and $ScO_6$ octahedra with W-O and Sc-O interatomic distances and bond angles are shown in Fig. 6. The $WO_6$ octahedra are highly distorted with O-W-O angles ranging from 72.73(14)-163.78(14)$°$ (cf. orthorhombic $Sc_2W_3O_{12}$: 106.5(8)-118.1(9)$°$ [46]). The smaller O-Sc-O angles range from 73.48(19) to 97.77(14)$°$ [cf. orthorhombic $Sc_2W_3O_{12}$: 84.0(5)-95.8(6)$°$]. The larger angles are on average more than 10 degrees smaller than those of the corresponding $Sc_2W_3O_{12}$ angles, suggesting that the $ScO_6$ octahedra are somewhat distorted. Comparison of the interatomic distances and angles with those of orthorhombic $Sc_2W_3O_{12}$ [8] shows that both the mean Sc-O and the W-O bond distances increase [from 2.096(12) to 2.193(7) Å, and from 1.756(16) to 1.920(6) Å, respectively], consistent with the change in the W coordination number from tetrahedral (WO$_4$) in Sc$_2$W$_3$O$_{12}$ to octahedral (WO$_6$) in Sc$_{0.67}$WO$_4$, and a slight distortion of the ScO$_6$ octahedra. The average W-O distance is similar to that found in MnWO$_4$: 1.937 Å. [24] Bond valence sum (BVS) calculations (soft valence [47]) result in BVS of 6.21 for W, and 2.36 for Sc. For comparison, the BVS of W in orthorhombic Sc$_2$W$_3$O$_{12}$ is 5.82, in MnWO$_4$ it is 5.94, while for Sc it is 2.88 in Sc$_2$W$_3$O$_{12}$. The deviation from the expected BVS in Sc$_{0.67}$WO$_4$ may be understood to result from the Sc vacancy structure. WO$_6$ octahedra proximate to Sc vacancies are expected to attract the now underbonded terminal O toward the W center, leading to an average d$_{W-O}$ shorter than a hypothetical defect-free structure. Likewise the average d$_{Sc-O}$ will be longer than expected for such a hypothetical structure. These defect-driven bonding effects lead to the calculated overbonded or underbonded BVS calculated for W and Sc, respectively.

Cation vacancies have been observed before in ABO$_4$ scheelites. For instance, Eu$_x$MoO$_4$ and Eu$_x$WO$_4$ (0.67≤x≤0) were shown to be mixed-valence phases containing both Eu(II) and Eu(III) ions, [43,48] with Eu being mostly Eu(II) in the vicinity of x=1 in Eu$_x$MoO$_4$. [49] Based on density measurements, Banks *et al.* suggested that the scheelite Eu$_x$MoO$_4$ solid solution series is characterized by Eu lattice vacancies. [43] Their IR spectroscopic and lattice parameter measurements indicated that the vacancies are compensated by the presence of Eu$^{3+}$, and at x=0.67, Eu$^{II}$Mo$^{VI}$O$_4$ changes to Eu$^{III}_{0.67}$Mo$^{VI}$O$_4$. The oxidation state of Mo in Eu$_x$MoO$_4$ was confirmed to be +6. [50] Our successful preparation of Sc$_{0.67}$WO$_4$ demonstrates that such A-site defects are also possible in the wolframite structure type.

Sleight has proposed a structure-sorting map for scheelites and wolframites based on unit cell volume of the ABO$_4$ phase and the cell edge of the corresponding AO rocksalt, [28] $Sc_{0.67}WO_4$ falls in expected range for wolframite type and indeed agrees well with the trends found in Fig. 1 of Reference 28. However, to our knowledge, $Sc_{0.67}WO_4$ is the first wolframite-type with a trivalent A cation. Its formation at high pressure is consistent with the general picture that the wolframites are higher density phases (with close-packed oxygens) often obtained by a reversible transformation from the scheelite structure on compression. [28] It should be noted that in his early paper, Sleight reported that he prepared a black $EuMoO_4$ phase, and speculated on the presence of $Eu^{3+}$ or $Mo^{5+}$ in the material. [28] Later, the $Eu_{0.67}MoO_4$ scheelite phase was reported by Banks et al., and the presence of $Eu^{3+}$ and $Mo^{6+}$ cations was established by IR spectroscopy and measurement of lattice parameters, as mentioned above. [43] The divalent scheelite $EuWO_4$, and mixed-valent scheelite phases with $Eu_xWO_4$ (0.8≤x<1.0) composition were prepared and their magnetic properties studied by Greedan et al. [48]

*4.2. Electrical and Magnetic Properties*

There have been several studies on the electronic structure of $ABO_4$ scheelites ($CaMoO_4$, $CaWO_4$, $CdMoO_4$, $PbMoO_4$, $PbWO_4$, [40,41,42,51] and $BaWO_4$ [42]) and wolframites ($CdWO_4$ [40] and $ZnWO_4$ [42]). The latter wolframites are wide-gap n-type semiconductors. [40,42,52] Their band gaps fall into the 2.4-6 eV range; e.g. 2.43 eV for $CdMoO_4$, [40] and 5.97 eV for $CaWO_4$. [42] Their valence and conduction bands are mainly composed of O 2p and W 5d states, respectively. There is some electronic contribution from Cd and Zn, mainly to the valence band, negligible to the conduction band. [40,42] In $CdWO_4$, W has a distorted octahedral coordination causing the lower part of the conduction band to be threefold degenerate. [40] Conductivity studies on related systems such as $AlWO_4$, [45] $FeMoO_4$, [53,54] $CoMoO_4$, [55] $NiMoO_4$, [56] and La-doped $PbWO_4$ [57] revealed activated behavior, polaron hopping, and oxide-ion conduction mechanisms with activation energies in the 0.2-2 eV range. [58] With its (poor) metallic conductivity and no evidence for activated behavior (*vide infra*), $Sc_{0.67}WO_4$ is clearly distinct from these related scheelites and wolframites.

Consideration of the magnetic susceptibility, thermoelectric power, and electrical conductivity indicate that $Sc_{0.67}WO_4$ is an n-type disordered metal. The modified Curie-Weiss fit to $\chi(T)$ reveals a TIP superimposed on a low-temperature Curie tail. Under the assumption that the local moments arise from localized S=1/2 sites, the extracted $p_{eff}$ = 0.15 $\mu_B$/f.u. puts these magnetic impurities at a concentration of ~0.8 mole percent. This is too large to come from impurities originating in starting materials or introduced during sample manipulation, and so we attribute this Curie behavior to a small fraction of localized $W^{5+}$ (S=1/2) sites in Sc-W-O sample. The origin of these local moments may be extrinsic (i.e. residing in impurity phases) or intrinsic to the $Sc_{0.67}WO_4$; our present data do not allow us to distinguish between these possibilities. Alternative explanations such as carrier 'freeze out' into localized states postulated recently in $Na_{0.3}RhO_2 \cdot 0.6H_2O$ are unlikely, [59] as the energy gap would have to be of order 2 K or less given the conductivity data of Fig. 5(b). The observed Pauli susceptibility reflects a finite density of states at the Fermi level, $E_f$. Using the free electron metal relationship $\chi_{Pauli} = \mu_B^2 N(E_f)$, we estimate $N(E_f)$ ~ 8 x $10^{24}$/eV•mol. An independent measure of $N(E_f)$ can be obtained from the specific heat (Fig. 4(b)), which at low temperatures is given by $C_p = \gamma T + \beta T^3$. The coefficient of the linear term, $\gamma$, is the electronic contribution, which in the free electron model is proportional to $N(E_f)$, i.e., $N(E_f) = 3\gamma/\pi^2 k_B^2$. Analysis of the data measured on S2 [60] shown in Fig. 4(b) yields $N(E_f) \sim 5 \times 10^{23}$/eV•mol. The discrepancy between the magnetic and thermodynamic measurements of $N(E_f)$ can be attributed to sample variation (S1 vs S2), sample aging, and/or imprecise correction of the measured magnetic susceptibility data; the first two possibilities are perhaps more likely considering sample variation of conductivity data shown in Fig. 5. Nonetheless, these measurements independently argue for a substantial, finite density of states at $E_f$ in $Sc_{0.67}WO_4$.

The finite density of states at $E_f$ is corroborated by the thermoelectric data of Fig. 5(a), which in the range 300 – 600 K follow the T-linear behavior expected for a metal. An unusual feature of $Sc_{0.67}WO_4$ is the electrical conductivity [Fig. 5(b)], which is approximately proportional to T over two decades with a finite T=0 intercept of ~0.08 S/cm, again arguing that $Sc_{0.67}WO_4$ is a (poor) metal. [61] Notably, there is no evidence for the activated behavior expected for semiconductors, polaronic metals, etc. Fig. 5(c) replots the conductivity data on logarithmic axes, revealing a low-temperature deviation from linearity. With the caveat that quantitative descriptions of conductivity in polycrystalline samples are inevitably subject to errors introduced by grain connectivity issues, the data can be well fit to a form appropriate for disordered metals, $\sigma = \sigma_0 + AT^{1/2} + BT$, [62] where the T and $T^{1/2}$ terms are characteristic of weak localization with a dominant dephasing by electron-electron scattering and Coulomb interaction terms, respectively. While this expression is strictly applicable to low-temperature conductivity, Helgren *et al*. [63] have successfully analyzed the conductivity of rare-earth and Y-doped Si up to 300 K using this expression. If such a model correctly describes the electrical conductivity in $Sc_{0.67}WO_4$, the disorder expressed by the Sc nonstoichiometry through its impact on the local structure of the $WO_6$ octahedra may be a relevant contributor to the localization mechanism.

Also shown in Fig. 5(c) are data from S2 as well as data collected on S1 following several months of storage in a nitrogen-purged storage cabinet. The absolute conductivity of the first sample has changed substantially during the storage period, but the data are still well-described by the same functional form as the initial data. We do not know why the conductivity of this sample has changed; possibly some reaction with the atmosphere has changed the carrier concentration and/or mobility. Alternatively an order-disorder process among defects in the structure may have occurred. Data from the second sample are clearly linear at T > 30 K, but the low T agreement is not as satisfactory. Additionally, the coefficient of $T^{1/2}$ is positive in the case of the first sample and negative for the second. These sample-related inconsistencies make it impossible for us to claim definitively that the disordered model is appropriate to $Sc_{0.67}WO_4$. Better samples and systematic studies will be necessary to answer this question. Regardless of the exact description of the conductivity mechanism, however, $Sc_{0.67}WO_4$ is metallic with a finite density of states at $E_f$; its conductivity is influenced by very strong disorder, and is perhaps not too different from heavily-doped amorphous systems. In an extensive literature search, we have found no other reports of metallic conductivity in doped scheelites or wolframites.

How are the carriers generated in this nominally $d^0$ compound? One potential mechanism for generating n-type carriers is through the introduction of oxygen vacancies, i.e., $Sc_{0.67}WO_{4-\delta}$. Many examples of such a mechanism are found in oxides: O vacancies are known to introduce carriers in $WO_{3-x}$, although the conduction mechanism in this compound is polaronic; [64] conductance through oxygen vacancy donor states is known to occur in the n-type semiconductor bismuth iron molybdate, $Bi_3FeMo_2O_{12}$; [65-67] oxygen vacancies lead to metals in reduced bulk $TiO_2$ Magneli phases, [68] and in thin films of $TiO_{2-x}$. [69] Notably, in all of the above examples, electron-lattice contributions are important to the conductivity; the same does not appear to be the true in $Sc_{0.67}WO_4$.

If we attribute the ~0.8% $W^{5+}$ concentration determined from magnetic susceptibility wholly to O defects in $Sc_{0.67}WO_4$ and assume each vacancy donates two free carriers to the band structure, this equates to an upper bound of 0.004 O vacancies per formula unit and a carrier concentration estimate of $n \sim 6 \times 10^{19}$ $cm^{-3}$. These vacancies can in principle form extended impurity states below and potentially merging with the conduction band. The measured IR feature at ~150 meV could reflect the residual presence of such impurity states. An alternative mechanism such as that proposed in $Bi_3FeMo_2O_{12}$; [65-67] in which conduction occurs in the impurity band at low temperature and in the conduction band at high temperature seems unlikely given the lack of any energy scale other than temperature in the conductivity data.

In spite of this potentially small concentration of vacancies, we attempted to measure δ in the proposed $Sc_{0.67}WO_{4-\delta}$ by oxidizing sample S1 on a TGA balance (Fig. 7). Surprisingly, heating the as-prepared sample in pure $O_2$ to 600 $^o$C led to a 1.6% mass *loss* without decomposition: X-ray diffraction on the end-product of the TGA treatment showed no significant change relative to the starting material (Fig. 7, inset), demonstrating the robust stability of $Sc_{0.67}WO_4$. In contrast, Banks reports that $Eu_{0.67}MoO_4$, upon grinding with KBr, readily converts to the monoclinic $Eu_2Mo_3O_{12}$ structure. [43] To rule out the possibility that the mass loss could arise from adsorbed moisture, we heated as-prepared $Sc_{0.67}WO_4$ in pure nitrogen to 400 $^{o}$C on the TGA balance. Assuming that any adsorbed moisture would be fully evolved by this temperature, an observed mass loss of less than 0.05%, indicates that any such moisture is only a small fraction of the 1.6% loss observed at 500 $^{o}$C in $O_2$. Curiously, in the temperature range 150 $^{o}$C to 400 $^{o}$C, the mass loss in $O_2$ exceeds that measured in $N_2$. At this time we cannot explain this unexpected behavior. Nonetheless, if ~1.6% mass loss in $O_2$ can be attributed solely to O vacancy formation, it equates to ~0.3 O per formula unit and indicates that $Sc_{0.67}WO_4$ is indeed prone to pronounced O defect formation. It is also conceivable that the configuration of the O vacancies (clustering, long-range order, etc.) changes over time, resulting in the aging effect observed in the electrical conductivity. Further detailed studies will be required to establish whether or not this oxygen defect model is appropriate to describe the chemical as well as physical properties of $Sc_{0.67}WO_4$.

## 5. Conclusion

We have prepared the new compound, $Sc_{0.67}WO_4$, from the NTE precursor material $Sc_2W_3O_{12}$ under high pressure, high temperature conditions. To our knowledge, $Sc_{0.67}WO_4$ is the first wolframite with a trivalent A-site cation. The bonding in $Sc_{0.67}WO_4$ can be understood by considering the defective Sc sublattice, while the magnetic and electrical properties are speculated to originate from a small concentration of oxygen vacancies, leading to $W^{5+}$ sites and band-like electron carriers that are strongly influenced by disorder. Thus, both structurally and electronically $Sc_{0.67}WO_4$ stands as a unique example among the wolframite class of oxides. More generally, the findings here encourage us that NTE materials can serve as precursors to other interesting new high-pressure phases with modification of bonding and electronic structure.

**Acknowledgments**

The work at Argonne National Laboratory, including the use of the Advanced Photon Source, was supported by the U.S. DOE Office of Science, Basic Energy Sciences, under Contract No. DE-AC02-06CH11357. Assistance by John Schlueter and Kylee Funk, and Maria Chondroudi (Materials Science Division, Argonne National Laboratory) with SQUID magnetic measurements is acknowledged. We also thank Prof. Mercouri G. Kanatzidis for providing access to the MMR (measurement of thermoelectric power) and FT-IR instruments. T.V. and J. F. M. thank Mark Bailey (Wildcat Discovery Technologies) for discussion of band structure calculations, and Qing'an Li and Kenneth Gray for measurement of and Konstantin Matveev and Chris Leighton (Univ. of Minnesota) for discussions of electrical conductivity data.

**Appendix A.   Supplementary material**

Supplementary data associated with this article can be found in the online version at doi:…

**Table 1 :** Lattice parameters, Sc/W fractional occupancies, and the amounts of crystalline phases present obtained from the Rietveld fits to synchrotron powder diffraction data from sample S1. Literature data for the model structure $MnWO_4$ are also included for comparison. [24]

| Measured formula | $a$ (Å) | $b$ (Å) | $c$ (Å) | $\beta$ (°) | $V$ (Å$^3$) | Sc occup. | W occup. | $Sc_{0.67}WO_4$ (wt%) | $Sc_2W_3O_{12}$ (wt%) | $Sc_{0.3}W_{0.85}O_3$ (wt%) |
|---|---|---|---|---|---|---|---|---|---|---|
| $Sc_{0.67}WO_4$ | 4.80282(4) | 5.75801(6) | 4.98611(5) | 91.177(1) | 137.860(2) | 0.669(4) | 1.0 | 86.96(10) | 3.55(6) | 9.49(6) |
| $MnWO_4$ | 4.830(1) | 5.7603(9) | 4.994(1) | 91.14(2) | 138.9207 | Mn : 1.0 | 1.0 | n/a | n/a | n/a |

**FIGURE CAPTIONS**

**Fig. 1:** Comparison of the x-ray diffraction patterns and appearance of orthorhombic $Sc_2W_3O_{12}$ (bottom pattern), and monoclinic $Sc_{0.67}WO_4$ (top). ×: residual orthorhombic $Sc_2W_3O_{12}$; ■: "$Sc_{0.3}W_{0.85}O_3$" impurity. Data collected using Co $K_\alpha$ radiation (0.78901 Å). Data are from sample S1.

**Fig. 2:** Typical Rietveld fit of the $Sc_{0.67}WO_4$ and its impurities. Data collected on sample S1 at a wavelength of 0.401555 Å. Tickmarks for the phases present are shown (refer to online version for colors): black - $Sc_{0.67}WO_4$; blue - $Sc_2W_3O_{12}$; green - $Sc_{0.3}W_{0.85}O_3$. $\chi^2$: 5.951, $wR_p$: 0.1158, $R(F^2)$: 0.0503.

**Fig. 3:** Structure of orthorhombic $Sc_2W_3O_{12}$ and monoclinic $MnWO_4$ represented by polyhedra. Dark octahedra: $ScO_6$ (left) and $MnO_6$ (right); light polyhedra: $WO_4$ (left) and $WO_6$ (right).

**Fig. 4:** (a) Temperature dependence of the magnetization of $Sc_{0.67}WO_4$ on cooling from 300 to 2 K. Applied field was 1 T. Inset: Data with low-temperature Curie tail subtracted showing temperature-independent Pauli component. (b) Specific heat of $Sc_{0.67}WO_4$ (Sample S2) plotted as $C_p/T$ vs $T^2$.

**Fig. 5:** (a) Temperature dependence of the thermoelectric power of $Sc_{0.67}WO_4$ (Sample S1) and the temperature dependence of electrical resistivity of $Sc_{0.67}WO_4$. (b) Temperature dependence of electrical conductivity and resistivity of $Sc_{0.67}WO_4$.

(c) Temperature dependence of electrical conductivity of $Sc_{0.67}WO_4$ samples with fits to models of disordered metal. S1: sample 1 as-made; S1-aged: sample 1 after aging; S2: sample 2. Values of fitting coefficients: S1 (initial measurement) : $\sigma_0 = 0.074$ S•cm$^{-1}$, A = 1.7 x 10$^{-3}$ S•cm$^{-1}$•K$^{-1/2}$; B = 1.0 x 10$^{-3}$ S•cm$^{-1}$•K$^{-1}$; S1-aged: $\sigma_0 = 0.029$ S•cm$^{-1}$, A = 3.4 x 10$^{-4}$ S•cm$^{-1}$•K$^{-1/2}$; B = 9.2 x 10$^{-4}$ S•cm$^{-1}$•K$^{-1}$; S2: $\sigma_0 = 0.014$ S•cm$^{-1}$, A = -2.6 x 10$^{-4}$ S•cm$^{-1}$•K$^{-1/2}$; B = 7.4 x 10$^{-4}$ S•cm$^{-1}$•K$^{-1}$ See text for details.

**Fig. 6:** Bonding environments of W and Sc determined from refinement of synchrotron x-ray powder diffraction data.

**Fig. 7:** Thermogravimetric data for $Sc_{0.67}WO_4$ (Sample S1) treated in oxygen (squares) or nitrogen (circles). Inset: Powder x-ray diffraction patterns of $Sc_{0.67}WO_4$ before and after thermogravimetric analysis in oxygen. The top trace has been offset by 15000 counts for clarity.

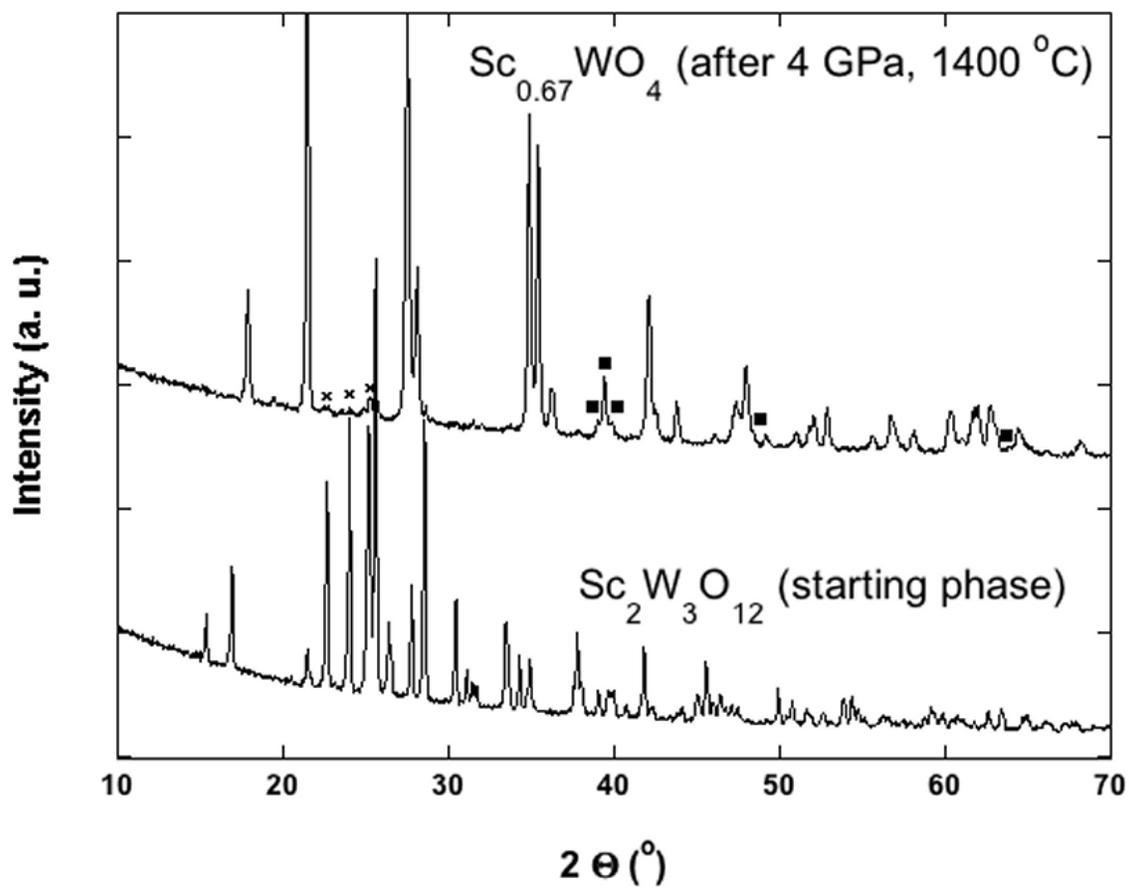

**Figure 1**

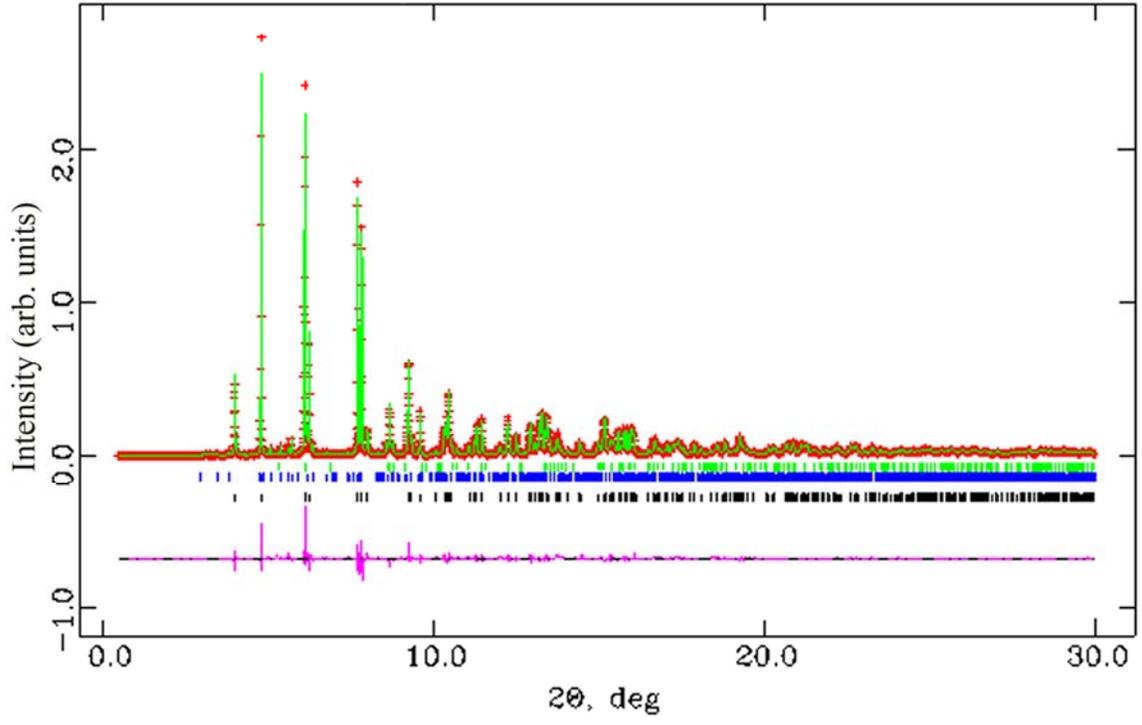

**Figure 2**

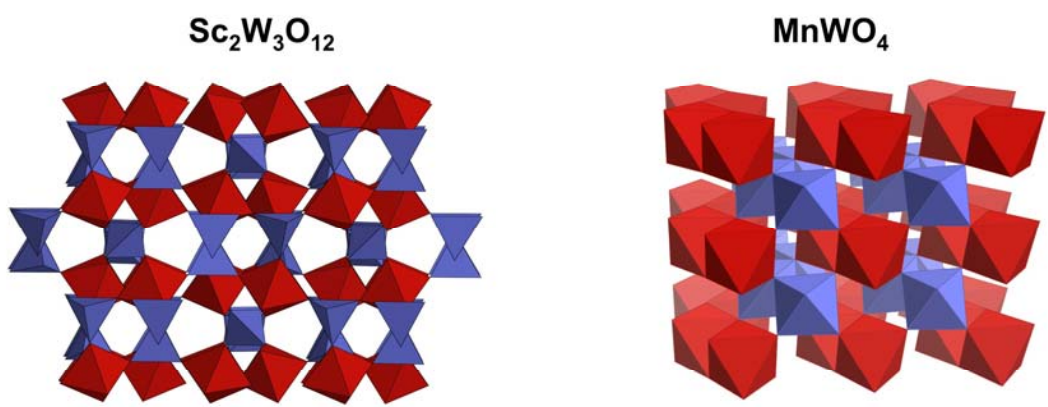

**Figure 3**

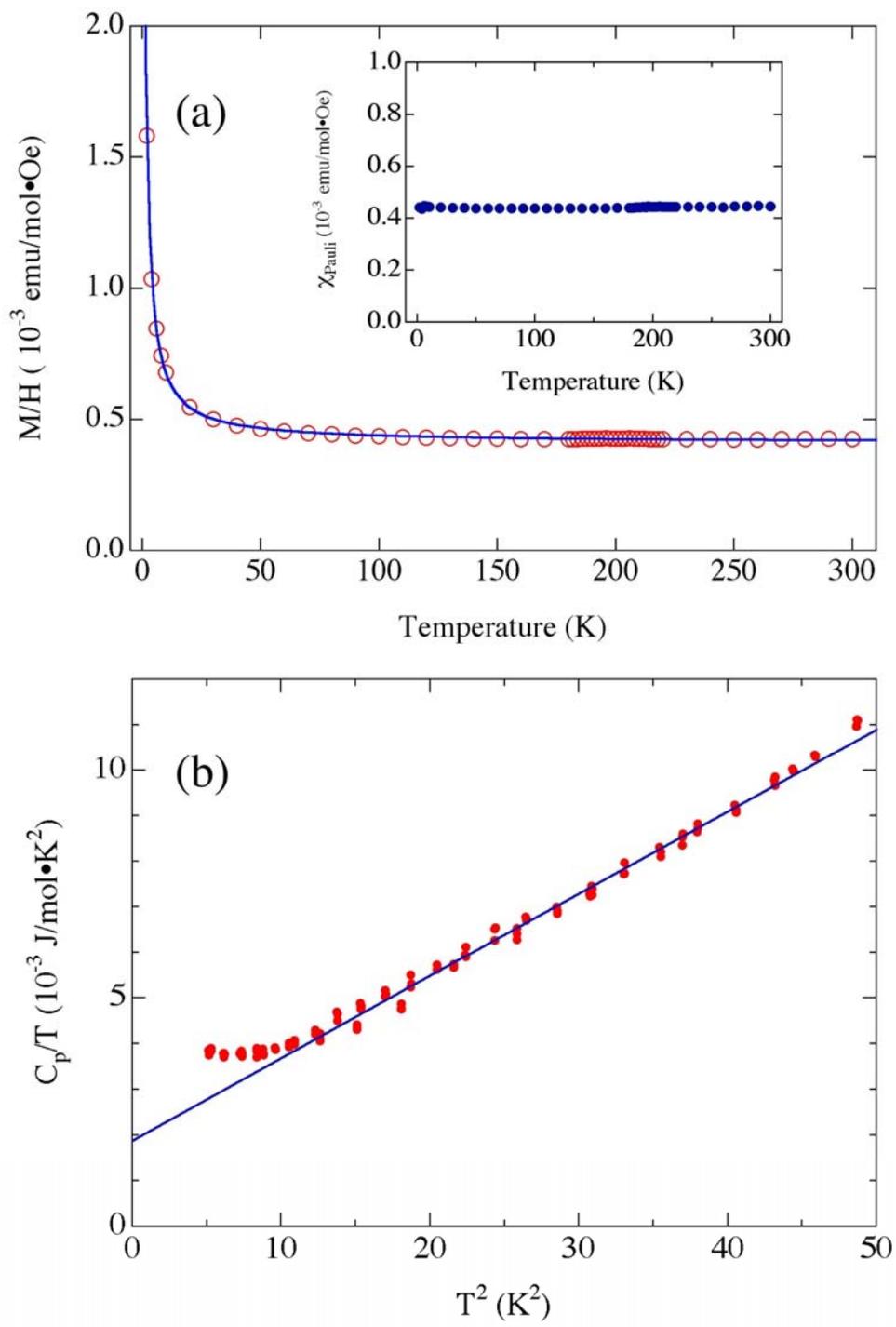

**Figure 4**

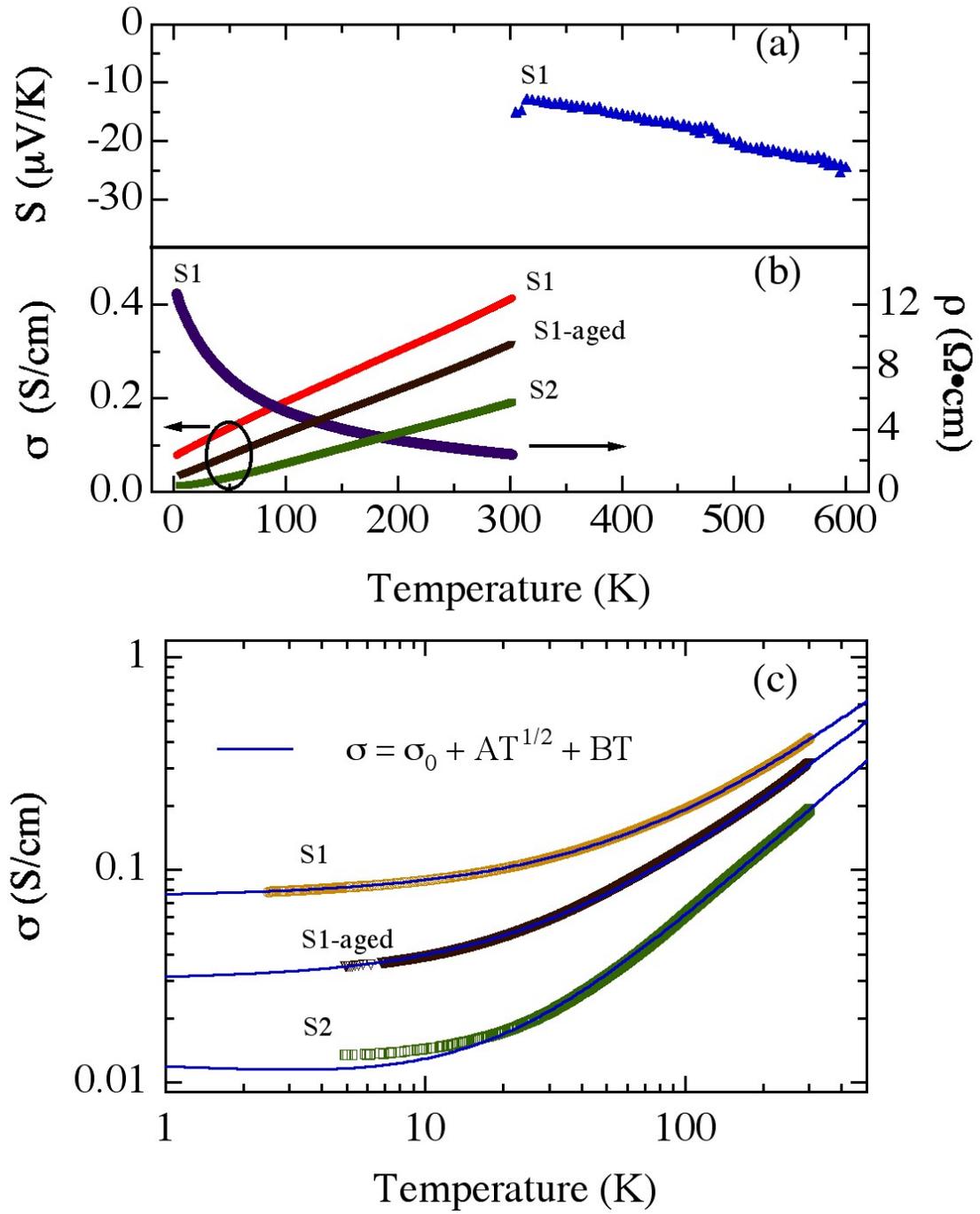

**Figure 5**

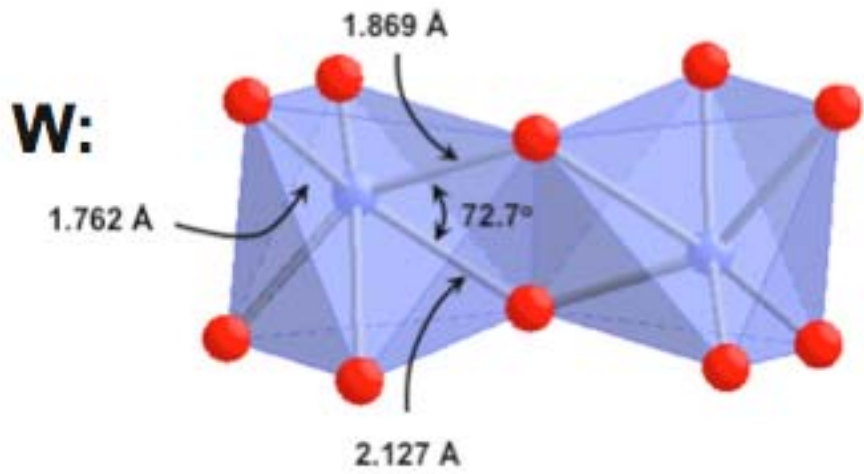
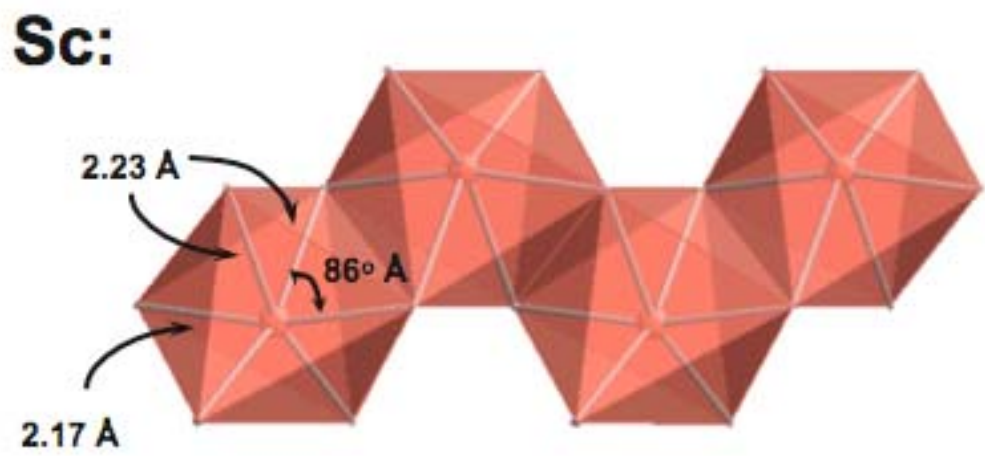

**Figure 6**

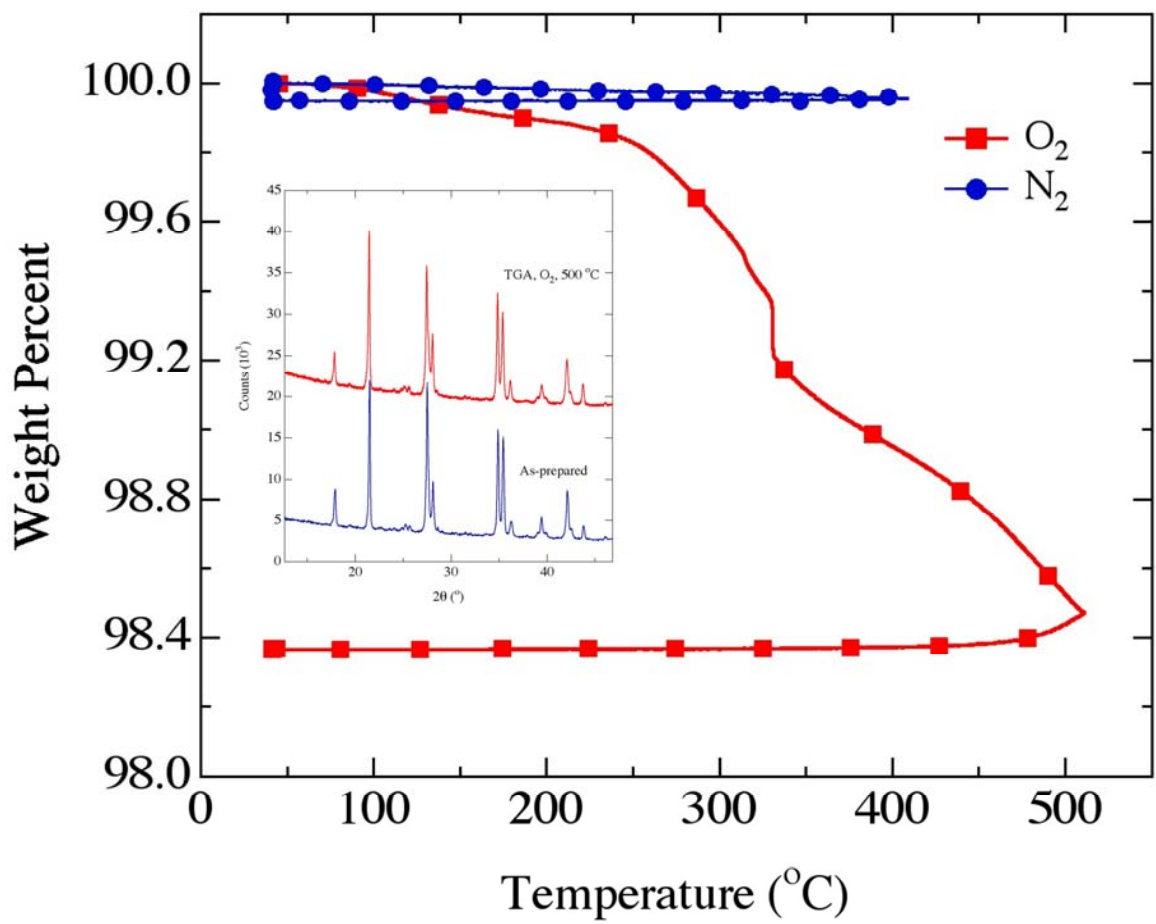

**Figure 7**